\documentclass{article}

\usepackage{arxiv}

\usepackage[utf8]{inputenc} 
\usepackage[T1]{fontenc}    
\usepackage{hyperref}       
\usepackage{url}            
\usepackage{booktabs}       
\usepackage{amsfonts}       
\usepackage{nicefrac}       
\usepackage{microtype}      
\usepackage{lipsum}		
\usepackage{graphicx}
\usepackage{natbib}
\usepackage{doi}

\title{Towards Dynamic Threat Modelling in 5G Core Networks Based on MITRE ATT\&CK}

\author{ 

\href{https://orcid.org/0000-0000-0000-0000}{\includegraphics[scale=0.06]{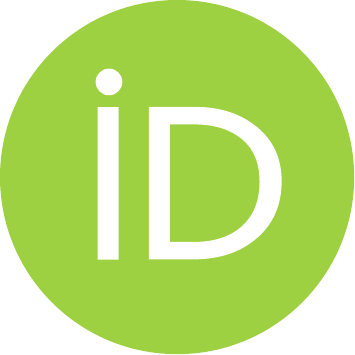}\hspace{1mm}Robert Pell}\\
	Department of Computer Science\\
	University of Surrey, UK\\ \\
	\texttt{r.pell@surrey.ac.uk} \\
	
	\And
	Sotiris Moschoyiannis\\
	Department of Computer Science\\
	University of Surrey, UK\\ \\
	\texttt{s.moschoyiannis@surrey.ac.uk} \\
	
	\And
	\href{https://orcid.org/0000-0001-7306-4062}{\includegraphics[scale=0.06]{orcid.pdf}\hspace{1mm}Emmanouil Panaousis} \\
	School of Computing and Mathematical Sciences\\
	University of Greenwich\\
	\texttt{e.panaousis@greenwich.ac.uk} \\
	
	\And
	\href{https://orcid.org/0000-0001-7306-4062}{\includegraphics[scale=0.06]{orcid.pdf}\hspace{1mm}Ryan Heartfield} \\
	School of Computing and Mathematical Sciences\\
	University of Greenwich\\
	\texttt{R.Heartfield@greenwich.ac.uk} \\
}

\renewcommand{\shorttitle}{\textit{arXiv} Template}

\begin{document}
\maketitle

\begin{abstract}
This article discusses how the gap between early 5G network threat assessments and an adversarial Tactics, Techniques, Procedures (TTPs) knowledge base for future use in the MITRE ATT\&CK threat modelling framework can be bridged. 
We identify knowledge gaps in the existing framework for key 5G technology enablers such as SDN, NFV, and 5G specific signalling protocols of the core network.
We adopt a preemptive approach to identifying adversarial techniques which can be used to launch attacks on the 5G core network (5GCN) and map these to its components. 
Using relevant 5G threat assessments along with industry reports, we study how the domain specific techniques can be employed by APTs in multi-stage attack scenarios based on historic telecommunication network attacks and motivation of APT groups. 
We emulate this mapping in a preemptive fashion to facilitate a rigorous cyber risk assessment, support intrusion detection, and design defences based on common APT TTPs in a 5GCN.
\end{abstract}


\section{Introduction}
Advanced Persistent Threats (APTs) are a formidable threat to the cybersecurity of modern technology ecosystems, routinely defeating sophisticated system security controls with Tactics, Techniques and Procedures (TTPs) often tailored to a target's environment and specifically designed to evade detection \cite{daly2009advanced,pitropakis2018enhanced}. 
In recent years, enabled by an acceleration in digital transformation and greater network connectivity, APT campaigns have proliferated to industries historically isolated from the Internet, such as critical national infrastructure and vendor supply chains \cite{lemay2018survey,ahmad2019strategically}. 
As a result, a growing democratisation and prevalence in APT-level capability among threat actors is leading to a paradigm shift in how cybersecurity practitioners and researchers approach threat modelling and risk assessments for security control selection of existing and emerging technological ecosystems.

The advent of 5G Core Networks (5GCN) promises to provide a vibrant and service-rich technology ecosystem converging distributed computing (Cloud, Edge, IoT), Software Defined Networking (SDN) and elastic application services capable of serving a wide range of state-of-the-art industry use cases in autonomous vehicles, industrial IoT and healthcare applications. 
Naturally, therefore, as the aperture of APTs continues to expand, 5GCN has been identified as an appealing target to threat actors such as APTs and to date numerous 5G network security assessments have been conducted by academics, security groups, and industry suppliers \cite{khan2019survey,ENISAThreatReport,geller20185g} in an attempt to identify suitable controls for protecting 5GCN architectures. 
However, the introduction of diverse new technologies and an elastic re-configurable architecture at the core of 5GCN's value proposition introduces new cyber challenges.  

Traditional approaches to threat modelling that focus on specific software vulnerabilities or a static network configuration and application services are likely to become quickly outdated or entirely redundant when: (i) the flexible 5GCN architecture changes; or (ii) when applied to a particular 5GCN instance where the software platforms and applications services, in use, are different or constantly evolving. 
This plasticity in the 5GCN introduces a particularly complex challenge for establishing standardised and analytically consistent processes for threat modelling, as the dynamic nature of 5GCN creates a malleable threat modelling landscape which evolves dynamically alongside the continuous reconfiguration of 5GCN itself. 
 
Well known threat modelling frameworks such as STRIDE \cite{shostack2014threat} and OCTAVE \cite{alberts2003introduction} typically formulate the modelling process by first defining a high-level abstraction of a target system, its sub-systems and interfaces for profiling potential system attackers, their methods and associated objectives. 
This definition provides a model of system security contexts that allows for the creation of a catalogue of possible threats to the system and selection of security controls which can be used to address them; based on the severity of the threat and the risk it poses to the system. 
In this manner, traditional threat modelling frameworks generate ``stationary'' representations of a system with predefined attacker profiles. 
This approach has practical limitations for effective and resilient threat modelling in 5GCNs.
This is because they are dynamic and heterogeneous by design with system security contexts, which adapt continuously based on evolving configuration of the 5GCN instance (e.g. multi-tenancy application services, network slicing, quality of service, elastic compute, and software-defined radio access). 

The MITRE ATT\&CK framework provides a foundation for flexible and dynamic threat modelling of malleable system environments like the 5GCN which bridge heterogeneous technologies (e.g. SDN, Cloud, and IoT). 
Through its unique identification and categorisation of adversarial Tactics, Techniques and Procedures (TTPs), it establishes an environment neutral threat modelling methodology that supports dynamic security control selection and reconfiguration based on the changing patterns of threat actor behaviour \cite{nisioti2021data}. 
As a result, the MITRE ATT\&CK threat modelling methodology operates independently of a system abstraction, such as the high-level description of a specific 5GCN architecture and its component interfaces, as the modelling process focuses on the determination of \emph{adversarial behaviours} of threat actors. 
Whilst this approach provides a solution to addressing the dynamic nature of the 5GCN, it is reliant on a knowledge base of TTPs already observed. 
Implementation of MITRE ATT\&CK will rely on the identification of 5GCN specific TTPs in a preemptive fashion.

In this paper, we propose how to extend the existing MITRE ATT\&CK framework to incorporate a set of 5G specific adversarial techniques and discuss their use in modelling threats. 
Our approach is vendor agnostic and can be applied to flexibly model threats to 5GCN environments independently of changes to the 5GCN configuration state. 
In summary, we contribute to the literature by:
\begin{itemize}
    \item applying MITRE ATT\&CK to systematically identify 5GCN infrastructure components at risk ``pre'' and ``post'' intrusion to support the threat modelling and risk assessment process of 5GCN environments.
    \item extending the MITRE ATT\&CK framework to include adversarial techniques that APTs can leverage to target and compromise 5GCN infrastructure and services; identifying techniques relating to key technologies such as SDN, NFV and Network Slicing which are not currently included in the existing TTP matrices.
    \item using existing threat intelligence surrounding the motivation and objectives of APTs, when targeting telecommunication networks and draw comparisons between historic attacks and how they may be realised in 5G networks utilising our identified techniques.
\end{itemize}

The remainder of the paper is organised as follows: 
Section \ref{sec:background} provides a background on the specific threats relating to 5GCN infrastructure and the suitability of MITRE ATT\&CK in relation to these. 
In Section \ref{sec:mitre5g}, we identify, based on current literature, a set of adversarial techniques which could be used to attack the 5GCN by APTs creating a 5G TTP knowledge base and map these to the 5GCN infrastructure components.
In Section \ref{sec:application}, we demonstrate the use of our proposed \emph{5GCN adversarial techniques} for integration into an expanded MITRE ATT\&CK framework to model multi-stage attack threats to the 5GCN.
Finally, Section \ref{sec:concl} concludes the paper providing a summary of the main contributions and highlighting future work to be undertaken to further develop the integration of 5GCN adversarial tactics and techniques with the MITRE ATT\&CK framework.

\section{Background}\label{sec:background}
\begin{figure}
\centering
    \includegraphics[width=1\textwidth]{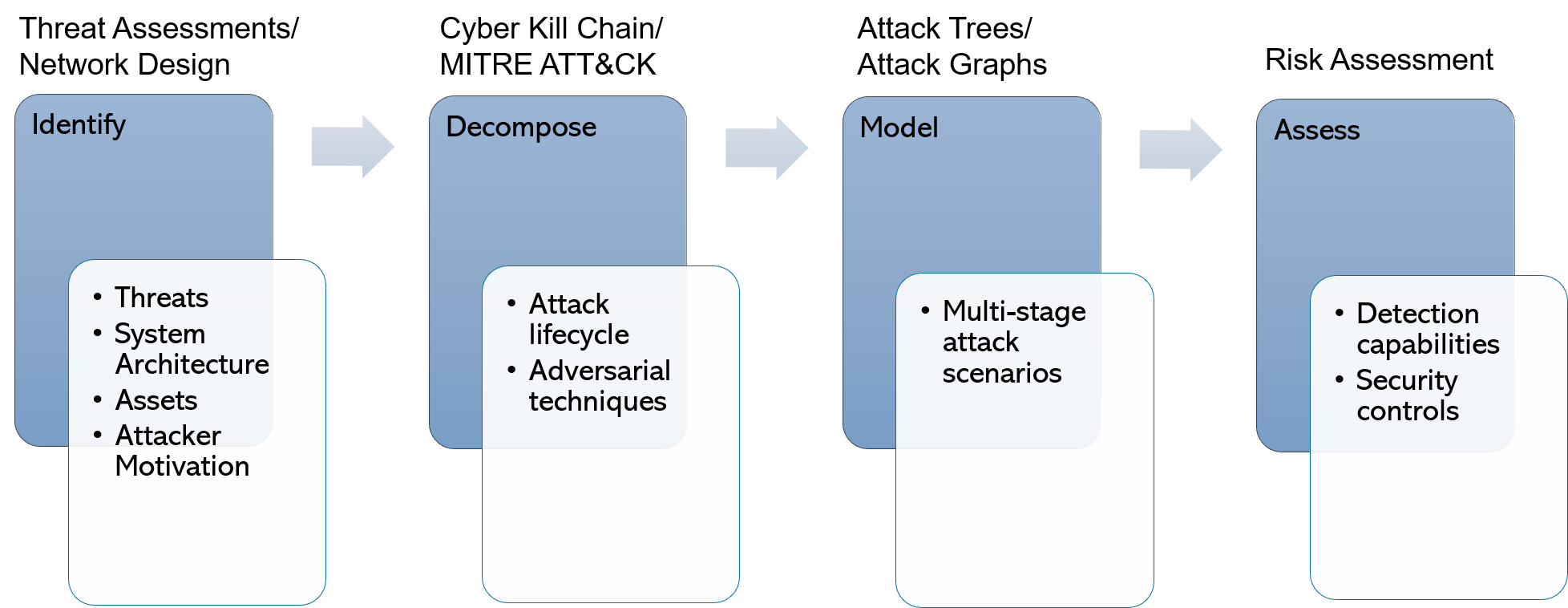}
	\caption[Security Assessment]{A Typical Process for Security Risk Assessment} 
	\label{Security Assessment Process}
\end{figure}

The attack surface of telecommunication networks is set to grow drastically with the introduction of new technologies and services to support new use cases in 5G networks. 
An important pre-requisite to the \textit{threat modelling} and \textit{risk assessment} processes is the identification of threats as shown in Figure \ref{Security Assessment Process}.
In the context of 5GCNs, effective threat modelling must consider an approach which divorces the modelling process from the need to define a static system abstraction for threat identification and instead fosters an approach which is independent of a specific 5GCN systems' configuration and architecture at a specific point in time.

MITRE ATT\&CK identifies a combination of techniques that an attacker may use.
These form specific attack scenarios expressed as attack graphs to demonstrate the multiple stages of an attack, without the dependency on defining a specific system abstraction. 
Unlike the traditional aforementioned high-level threat modelling methods, where selection of appropriate security controls are based on high-level descriptions of threats related to abstracted system security contexts, for each adversarial technique in the ATT\&CK knowledge database there exists specific \textit{detection} and \textit{mitigation} techniques.
These can be used to address applicable threats, regardless of how the particular environment, under investigation, is configured. 
This approach is considered a mid-level abstraction approach to modelling \cite{strom2018mitre} when compared to high-level abstraction frameworks such as the STRIDE and OCTAVE, or low-level exploit and vulnerability threat modelling, such as CVSS \cite{mell2006common}.

As the MITRE ATT\&CK framework is formulated by references to specific adversarial behaviour, its formulation and expansion largely follows observations drawn from APT campaigns and attack vectors observed in the wild. 
As a result, it can be considered a proactive event-driven framework which is updated as new adversarial behaviours are observed, analysed and systematically recorded. 
After the inception of a MITRE ATT\&CK TTP matrix for Enterprise Networks, MITRE extended the framework to include an Industrial Control System (ICS) network and mobile platforms, as Tactics and Techniques have been discovered from observations of attacks against these systems. 
Whilst 5G networks are relatively new and therefore ``5G specific'' attack vectors are yet to become commonplace, technologies and practices utilised in 5GCN infrastructure and services inherit many overlapping properties associated with both traditional enterprise networks and the cyber-physical aspects of ICS networks. 
Therefore a 5GCN knowledge base can benefit from an established baseline of known adversarial TTPs which apply to them for modelling adversarial threats against 5GCN environments. 
However, naturally, in MITRE ATT\&CK there currently exists a knowledge gap of potential future adversarial techniques which may be used to target components of 5GCN such as Software-Defined Networks (SDN), Network Function Virtualisation (NFV) and distributed cloud architecture, which are yet to be addressed by MITRE ATT\&CK. 

In this paper, we study how to extend MITRE ATT\&CK TTP knowledge base for 5G networks. 
The authors of \cite{rao2020threat} propose the \textit{Bahdra framework} which is a domain specific threat modelling framework for telecommunication networks. 
The framework organises the attack life-cycle into 3 stages consisting of 8 tactical groups of 47 techniques aligned to ATT\&CK. 
The main key differences between this work and ours are the following. 

First, the framework omits some tactics included in the ATT\&CK framework. 
A significant exclusion is that of data exfiltration which we believe is still a prime motivation of APTs when targeting 5G networks. 
In our work we include all of the tactic groups defined in ATT\&CK as we believe that the transition towards a traditional cloud computing architecture mean that 5G network inherit the same threat vectors in addition to the 5G specific ones. 
The 5G specific TTPs we identify are an extension of the existing MITRE knowledge base for cloud computing and network threats. 

Moreover, \textit{Bahdra framework} includes a new tactic Standard Protocol Misuse. 
Whilst we also include the misuse of protocols in a TTP knowledge base for 5G networks, we have included them as techniques which can be used to achieve some attack goals and not as a standalone tactic and therefore belong to the existing set of tactics in the ATT\&CK framework.
An example of this is the misuse of signalling protocols to evade network defences by blending into standard traffic. 

Furthermore, the proposed framework does not include techniques which span multiple tactic groups. 
In our work here, we identify some techniques which can be used for multiple tactics such as the use of standard network protocols for evading defences as well as for establishing Command and Control (C2) channels for data exfiltration. 
This factor can have an impact on assessing security risk to the network, if a specific technique can be used to achieve multiple tactical stages of an attack, it may be given greater weighting when considering security risk, mitigating actions and security controls.
Last, our work includes techniques which apply to the new 5G techniques such as SDN, NFV and network slicing which neither \textit{Bahdra} nor ATT\&CK include in their knowledge bases.

\subsection{5G Threats}
\begin{figure}
\centering
    \includegraphics[width=0.98\textwidth]{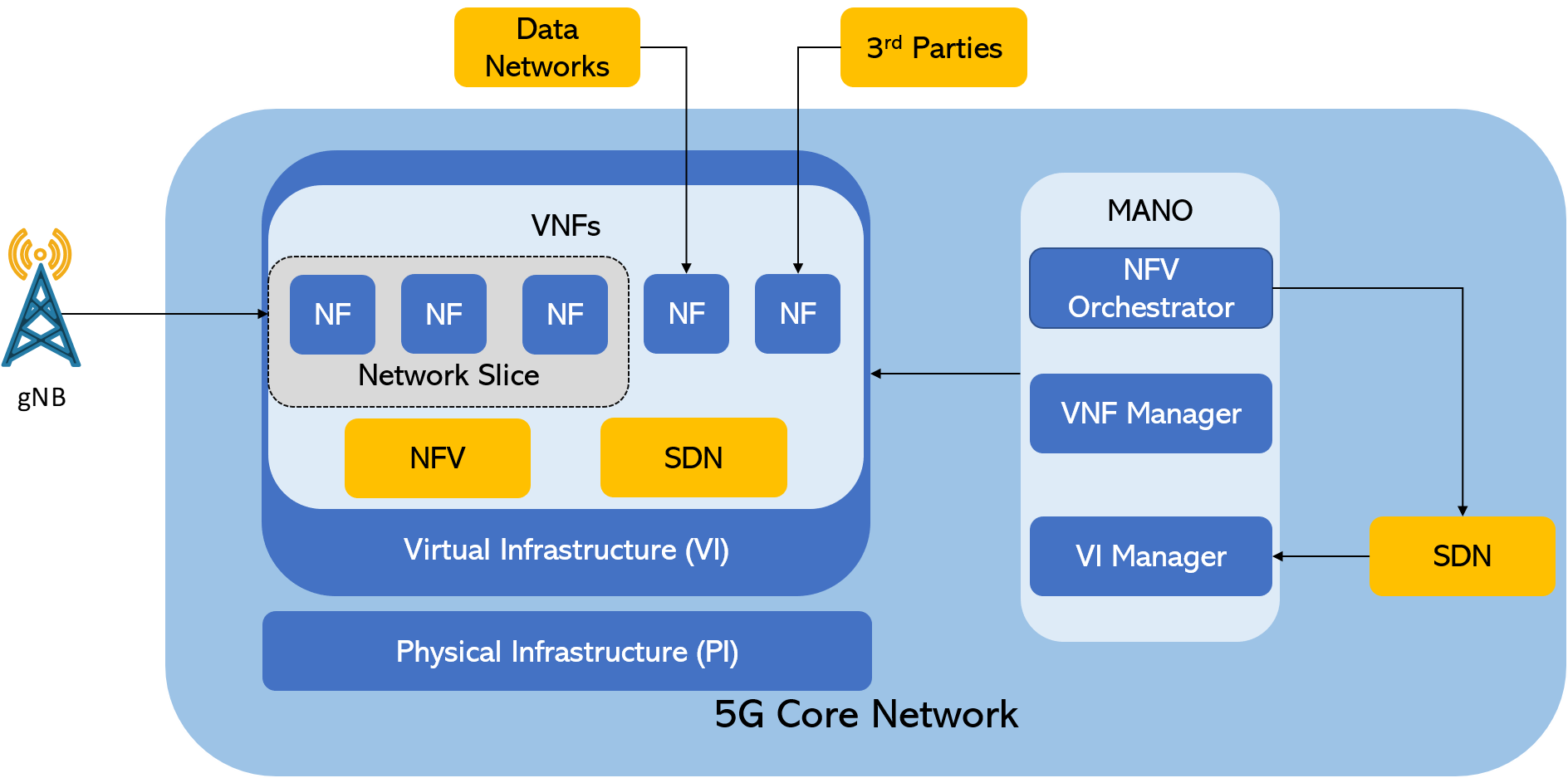}
	\caption[Scenario]{The 5GCN Architecture} 
	\label{5GCNArchitecture}
\end{figure}

We have identified, through literature review, threat scenarios which relate to the new 5GCN architecture and technologies used. 
Fig. \ref{5GCNArchitecture} provides a generic reference architecture of the 5GCN. 
The components are arranged based on the infrastructure layers and functional role within the 5G network architecture. 

The reference diagram provides an overview of the enabling technologies and the interfaces between the building blocks of 5G networks.   
The 5G network functions are implemented at the virtual layer through Network Function Virtualisation (NFV) and represent the application layer which can be further divided into the control plane and data plane.
Management and Orchestration (MANO) services are provided at multiple layers of the network architecture and SDN is utilised to provide a programmable interface for network management.

The shift towards a cloud based service deployment means that remote services such as the MANO provide a new attack vector which could be exploited \cite{ahmad2018overview}. 
Given the dynamic and re-configurable nature of the 5GCN for service provision on demand, configuration changes may be common and detection of malicious or erroneous actions becomes challenging. 
Further, there will be a heavy reliance on NFV and SDN in 5G networks to support new services which introduces new attack vectors. 
Threats to the SDN technology include Denial of Service (DoS) attacks and Man in the Middle (MiTM) attacks to modify network flow rules \cite{shu2016security}. 
The SDN flow tables provide valuable information about traffic routing within the core network and may be the target of discovery techniques as an attacker looks to identify target assets and navigate the network post intrusion. 
Prior to 5G, telecommunication networks contained many physical hardware devices responsible for providing service specific functionality in the core but these will now be virtualised service instances in the core network through NFV. 
New attack vectors associated with NFV include isolation failure of virtualised components and targeting the VNF components with DoS attacks to exhaust underlying shared resources \cite{lal2017nfv}.

Quick and dynamic VNF deployment also leads to the potential of configuration errors which introduce vulnerabilities within the network.
Besides the virtualisation aspect of the Network Functions (NFs) within Service Based Architecture (SBA) of the 5GCN, threats which target the Control Plane Signalling (CP) are of concern. 
Each NF can take on the role of the service provider or service consumer, which is managed by the Network Repository Function (NRF). 
Some NFs also interface with external networks such as the Access Management Function (AMF), Network Exposure Function (NEF), Security Edge Protection Proxy (SEPP) and User Plane Function (UPF). 
Other services provided by NFs of the SBA include the Authentication Server Function (AUSF), Session Management Function (SMF), Policy Control Function (PCF) and Network Slice Selection Function (NSSF). The Unified Data Management (UDM), Unified Data Repository (UDR) and Unstructured Data Storage Function (USDF) all serve as data repositories for providing data storage and access to the various NFs.


With core NFs becoming exposed to external networks and offering common interfaces by adopting an IP based protocol stack, there is a risk they could become targeted by attackers well versed in these technologies. 
Compromised NFs within the 5G core could lead to unauthorised access to data repositories, eavesdropping SBA communication, malware distribution, MiTM and DoS attacks \cite{rudolph2019security}. 

Service abuse is another motive for attacks on 5G networks. 
In a roaming scenario, the home network interfaces with a third party serving network that introduces the possibility of service abuse such as service fraud or provide a platform for DoS attacks through a compromised or insecure trusted mobile network operator. 
The lawful interception function provides emergency services the authorisation to intercept mobile communications for lawful purposes which could be used to access communications and confidential subscriber data if accessed by an adversary.

\section{MITRE ATT\&CK in 5G}\label{sec:mitre5g}
The MITRE ATT\&CK framework provides a TTP Matrix for common enterprise and Industrial Control System (ICS) network types as well as a matrix for mobile devices.
The matrices can easily be navigated through filtering of techniques based on network properties or technologies including host operating system, cloud based deployments, and network infrastructure. 

In order to apply the ATT\&CK framework for threat modelling in the 5GCN, an extension of the framework is required to include adversarial techniques relating to the 5GCN threats. 
Many of these techniques relate to the technologies used such as SDN, NFV and network slicing, and the SBA which are not currently contained within the ATT\&CK knowledge base. 
Table \ref{tab:techniques} provides an overview of TTPs relating to network technologies and assets based on the existing TTP matrices and those which, we believe, must be included in a 5G specific one.
Whilst the MITRE TTPs are a knowledge base of known APT behaviours, we aim to take a proactive approach to adversarial technique identification targeting these infrastructure components. 

\begin{table}[!b]
\caption{Comparison of Existing Frameworks}
{\begin{tabular*}{\textwidth}{@{\extracolsep{\fill}}lllll@{}}\toprule
\textbf{Technology Aspects} & \textbf{Enterprise} & \textbf{Mobile} & \textbf{ICS} & \textbf{5G Core} \\
\midrule
      Network & \checkmark & \checkmark & \checkmark & \checkmark \\ 
      OS & \checkmark & \checkmark & \checkmark & \checkmark \\ 
      Cloud Infrastructure & \checkmark &  &  & \checkmark \\ 
      Virtualisation/Containerisation & \checkmark &  &  & \checkmark \\ 
      Cyber Physical &  &  & \checkmark & \checkmark \\ 
      Industry Specific Protocols & \checkmark & \checkmark & \checkmark & \checkmark \\ 
      SDN &  &  &  & \checkmark \\ 
      NFV &  &  &  & \checkmark \\ 
      Network Slicing &  &  &  & \checkmark \\ 
      MANO &  &  &  & \checkmark \\ 
      5G Procedures &  & \checkmark &  & \checkmark \\ 
\bottomrule
\end{tabular*}}{}
\label{tab:techniques}
\end{table}

\subsection{5GCN Adversarial TTP Identification}
The MITRE ATT\&CK TTP matrices are based on \textit{threat intelligence} such as incident reports based on known \textit{APT campaigns}. 
MITRE identifies eight APT groups attributed with attacks on telecommunication networks (APT19, APT39, APT41, Deep Panda, MuddyWater, OilRig, Soft Cell and Thrip). 

\begin{figure*}[!t]
\centering
\includegraphics[width=1\textwidth]{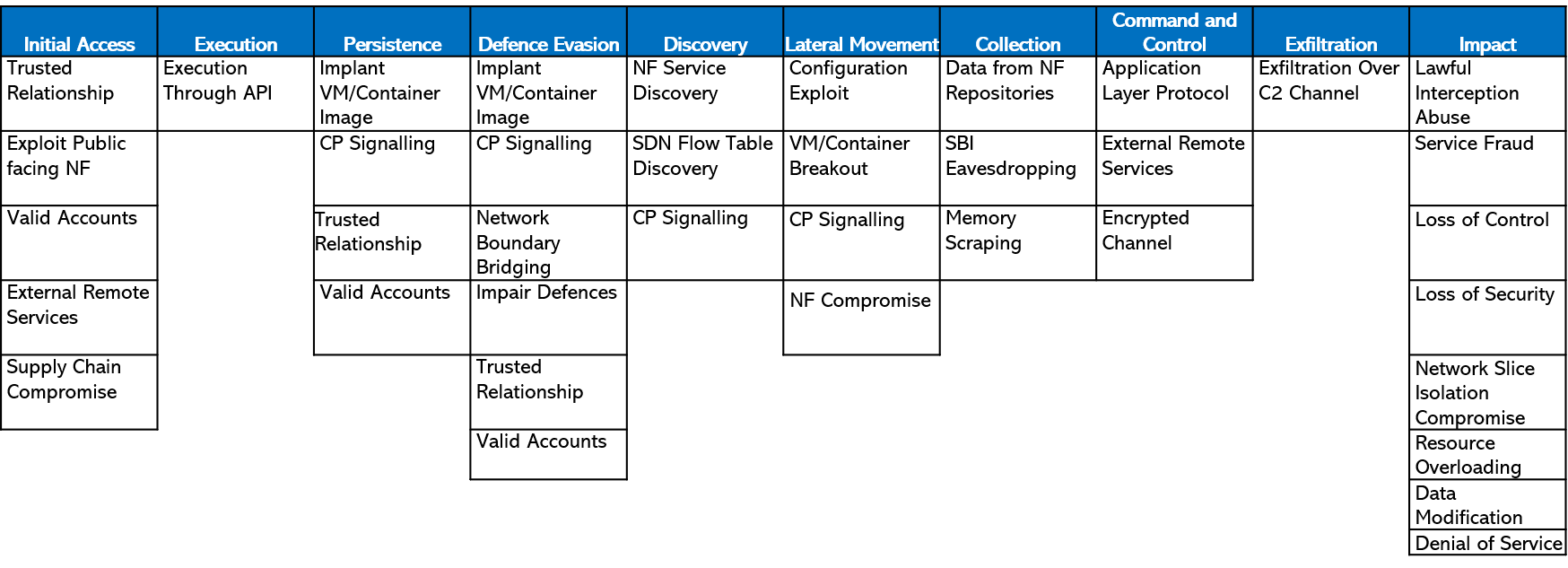}
\caption{5G APT TTPs}\label{fig:CombinedTTPs}
\end{figure*}

Given that 5G networks are an amalgamation of existing and new technologies, a 5GCN TTP knowledge base should extend the existing TTP matrix for the underlying cloud infrastructure and network but also include new techniques to incorporate new technologies such as SDN, NFV, and network slicing. 
Figure \ref{fig:CombinedTTPs} shows the proposed 5GCN TTPs which form the extended TTP knowledge base when combined with the existing knowledge base of TTPs used by APTs who have historically targeted telecommunication networks.

The techniques belonging to the MITRE ATT\&CK TTP matrices are generally abstract definitions of attack steps.
In some cases there exist sub-techniques which provide finer granularity offering more specific information about a particular technique.
The newly identified 5GCN techniques are either not included in the existing TTP matrices or are sub-techniques which have specific context in 5G networks.
For each tactic of the attack life-cycle we identify and characterise new techniques and explain the rationale behind their inclusion in the 5G TTP knowledge base.

\subsubsection{Pre-Intrusion \textit{(Initial Access, Execution)}}
We refer to the combination of Initial Access and Execution tactics as those of pre-intrusion, i.e. those which an attacker can use to gain access to the target network. 
In 5G networks we identify several access points which could serve as attack vectors for network intrusion.
Within the core network, NFs adopt common protocols and become accessible via RESTful APIs. 
The APIs exposed to external networks, such as that of the NEF and UPF provide a new attack vector to the 5GCN, such as those identified by the Open Web Application Security Project (OWASP) \cite{van2017owasp} and security concerns over the use of REST APIs \cite{yarygina2018overcoming}.
Trust relationships within 5G networks introduce significant security challenges \cite{rao2020threat}. 
This includes trusted relationships between home and visited networks in the roaming scenario, distributed network components across MEC deployments, third party applications and services, and between NFs belonging to different network slices. 
Attack scenarios where trusted relationships with roaming partners and between network slices are exploited have already been identified \cite{3GPP5GSecurityStudy, AdaptiveMobileSecurity}.
External remote services such as those provided by the MANO component provide important functions to network operators but also introduce attack vectors to provide access to the 5G network such as through the use of valid user accounts.

\subsubsection{Post-Intrusion \textit{(Persistence, Defence Evasion, Discovery, Lateral Movement, Collection, C2)}}
Post-intrusion tactics relate to those techniques used by an adversary following the initial intrusion and prior to the final objective. 
These activities can be thought of as those which help position the attacker to carry out the final stage of an attack whilst remaining undetected. 
In the context of APTs, this is inclusive of activities to help the attacker establish a foothold, evade defences, remain persistent, and move laterally to the target within the network. 
The use of virtualisation and container technology for the orchestration of NFs within the 5GCN introduce the possibility of image files being implanted within the system as to introduce a malicious NF \cite{sultan2019container},\cite{gao2017containerleaks}. 
This type of technique could arise directly as a result of a supply chain compromise or be facilitated by a malicious insider. 
This type of technique serves as a form of Persistence and Defence evasion tactic. 
The use of virtualised components which reside on the same physical infrastructure also introduces the risk of VM/container breakout, allowing an attacker to move laterally if there exists vulnerabilities in the hypervisor software or mis-configurations such as poor isolation of virtual machines and containers.

Trusted relationships, as with the Initial Access tactic, relate to the vulnerabilities introduced by trusted relationships. 
In the context of post intrusion techniques this refers to the trusted relationships within the 5GCN such as that between of NFs within the SBA and network slices. 
Once a NF has been authenticated and authorised within the SBA there is no guarantee these assets cannot be compromised and turned against the network to abuse the trust relationship through signalling abuse \cite{AdaptiveMobileSecurity, 3GPP5GSecurityStudy}. 
Many of the security controls within the SBA work at the network/transport layer but this does not protect against application layer misuse or abuse. 
These types of techniques can be used to evade defences and remain persistent given the lack of application level security and complexity of maintaining clear trust boundaries in a dynamic environment. 
CP signalling or rather misuse of it, is a sub-technique of the generalised standard protocol misuse technique set. 
This type of adversarial technique has been observed in attacks on telecommunication networks  including those which target legacy protocols such as SS7, Diameter and GTP. 
CP signalling is based on a set of HTTP messages defined in the 3GPP standard and therefore adversaries are well aware of the type of signalling used within 5G networks. 
The impact of CP signalling abuse is wide ranging from DoS, data leakage, service fraud, and data integrity security risks. 
Due to the fact that this technique is misuse of a valid protocol, it can be very difficult to detect malicious behaviour and as such can serve as a technique for achieving Persistence, Defence Evasion, Discovery, Lateral Movement, Data Collection and C2 tactics. 

A significant concern to network security should be the fact that CP signalling misuse can be deployed in a wide range of tactical steps.
We identify the potential for an adversary to gain initial access through the exploitation of the remote service offered for network MANO. 
Considering this as the first step in an attack, there arises the possibility of misusing the service to make configuration changes to the 5GCN \cite{homoliak2019insight}. 
Potential changes to the configuration can lead to changes in SDN flow tables, network configuration to impair defences or result in network boundary bridging. 
A more general purpose of this type of technique may be to introduce network configurations that can be exploited for lateral movement purposes whilst disguising configuration changes as intentional. 

In all types of networks, data collection poses a significant risk. 
Whether it be for the purpose of extracting data from the network or for gaining useful information about the network to further attacks, data collection techniques present a significant challenge to security. 
We have identified collection techniques which range from accessing data directly from NF repositories, which include crafting HTTP requests or exploitation of the databases themselves, memory scraping which targets the underlying memory in the cloud infrastructure, and passive attacks such as eavesdropping the Service Based Interface (SBI).  

Command and Control (C2) tactics serve the purpose of providing an attacker with a covert channel for their activities. 
This could be for providing a way of exfiltrating sensitive data outside of the network, maintaining a backdoor connection to the network or generally disguising communication between the target network and the outside world. 
In 5G networks, examples include use of applications layer protocols to disguise malicious communication between roaming partners, MEC services or third party applications. 
We have already identified the NF components, which provide an interface to the 5GCN, as a potential initial access point which logically could serve as a form of C2 channel if compromised. 
As with NF compromise the legitimate external services to MANO components could equally serve as a C2 channel.

\subsubsection{Objectives \textit{(Exfiltration, Impact)}}
Objectives reflect the goal of attacks and the ATT\&CK framework defines this as either Data Exfiltration or Impact techniques. 
Our contribution to extending the existing TTP knowledge base is the addition of Impact techniques, which are most relevant to 5G networks and have been identified in literature as potential threats. 
Impact techniques may be either combined for fulfilling the adversaries' end goal or adopted in a standalone manner.
Resource overloading and network slice compromise techniques are the result of some prior actions but can also be leveraged to advance attacks such as to induce a DoS or cause data leakage from a network slice. 
Data Modification can have wide a ranging impact on the network from corrupting data repositories to interfering with and adversely affecting cyber-physical systems.
Service specific impacts include the Abuse of the Lawful Intercept Function and Service Fraud which can be realised through charging and billing fraud.
Loss of security or control over parts of the network can have a significant impact on incident response and recovery depending on the level of control the adversary has achieved. 

\begin{table*}[!b]
\caption{Mapping of Adversarial Techniques to 5G Components}
{\begin{tabular*}{\textwidth}{@{\extracolsep{\fill}}lllllll@{}}\toprule
     \textbf{5G Technique} & \textbf{Physical} & \textbf{Virtual} & \textbf{NF} & \textbf{SDN} & \textbf{MANO} & \textbf{Network Slice} \\
\midrule
     Valid Accounts & & & & & \checkmark & \\
     Exploit Public Facing NF & & & \checkmark & & &\\
     External Remote Services & & & & & \checkmark &\\
     Supply Chain Compromise & & \checkmark & \checkmark & \checkmark & \checkmark &\\
     Execution through API & & & \checkmark & & \checkmark &\\
     Implant Container/VM Image & & \checkmark & \checkmark & & &\\
     Network Boundary Bridging& & \checkmark & \checkmark & & & \checkmark \\
     CP Signalling & & & \checkmark & \checkmark  & &\\
     Impair Defences & \checkmark & \checkmark & \checkmark & \checkmark & & \checkmark\\
     NF Service Discovery & & & \checkmark & & &\\
     SDN Flow Table Discovery & & & & \checkmark & &\\
     Configuration Exploit & \checkmark & \checkmark & \checkmark & \checkmark & \checkmark & \checkmark\\
     Container/VM Breakout & & \checkmark & \checkmark & & &\\
     NF Compromise & & \checkmark & \checkmark & & &\\
     Data from NF Repositories & & & \checkmark & & &\\
     SBI Eavesdropping & & \checkmark & \checkmark & & &\checkmark \\
     Memory Scraping & \checkmark & & & & &\\
     Application Layer Protocol (C2) & & \checkmark & \checkmark & & &\\
     External Remote Services (C2) & & & & & \checkmark &\\
     Encrypted Channel (C2) & & \checkmark & \checkmark & & &\\
     Exfiltration over C2 & & & & & \checkmark &\\
     Service Fraud & & & \checkmark & & &\\
     Loss of Control & \checkmark & \checkmark & \checkmark & \checkmark & \checkmark & \checkmark\\
     Loss of Security & \checkmark & \checkmark & \checkmark & \checkmark & \checkmark & \checkmark\\
     Network Slice Isolation Compromise & & & & & & \checkmark\\
     Resource Overloading & \checkmark & \checkmark & & & &\\
     Data Modification & & \checkmark & \checkmark & \checkmark & \checkmark & \checkmark\\
     Denial of Service & \checkmark & \checkmark & \checkmark & \checkmark & \checkmark & \checkmark\\
     \bottomrule
\end{tabular*}}{}
\label{tab:mapping}
\end{table*}

\subsection{Mapping Adversarial Techniques to 5GCN Infrastructure}\label{sec:MappingTTPs}
MITRE ATT\&CK positions itself as a mid level abstraction of adversarial threat modelling. 
Whilst it provides finer granularity and more details that high level abstraction frameworks, it lacks the low level information about the use of adversarial techniques. 
As a result of this, focusing resources to provide suitable security monitoring for detection and applying security controls to mitigate threats requires an understanding of which network assets are vulnerable to different techniques. 
Furthermore once adversarial techniques from the 5GCN knowledge base are assigned to the ``at risk'' assets, it can support threat modelling of multi-stage attacks through identification of paths an intruder could take. 
This can be represented through the use of attack graphs and used to support risk assessments by analysing each possible attack path.  

To achieve this we utilise the mid-level abstraction approach of MITRE and correlate techniques to the technologies of the 5G network architecture. 
For example; the exploitation of APIs technique is assigned to NF components of 5G networks rather than a specific instance of a NF such as the NEF. 
The reason for this approach is to firstly help focus security analysis to the assets which come under a technology category.
Second, 5G networks are dynamic, have variations in deployment, and given that the specification is still under development, there may still changes to the general network architecture. 
If a new NF is deployed, it will have the same security requirements as all other NFs in the network.
For this reason our approach to threat modelling can be applied to asset types regardless of network topology but still addresses threats to each technology group such as SDN, virtualisation, and asset properties such as trust relationships or protocol usage.  

Table \ref{tab:mapping} presents the mapping between the core infrastructure components and the 5GCN adversarial TTPs. 
This shows which of the 5GCN infrastructure components are at risk or impacted, in the case of adversary goals, of each adversarial technique we have identified.
General techniques or those that target windows operating systems or standard enterprise applications are not included as our focus is the mapping between the techniques and specific 5GCN components.
In some instances there may be multiple ``at risk'' components. 
For example SBI eavesdropping can occur through a compromise NF, NFs within the same network slice or through monitoring of the virtual network traffic thus applicable to the virtual, NF and network slice assets.

\section{Multi-Stage Attack Modelling with 5GCN TTPs}\label{sec:application}
So far we have identified a set of 5GCN adversarial techniques and created a mapping between those and the 5GCN infrastructure components. This equates to a 5GCN knowledge base of adversarial TTPs applicable to the network components and technologies. MITRE ATT\&CK is a threat modelling framework which provides TTPs as an input to the modelling process. 

In this section we demonstrate how the matrix consisting of 5GCN techniques can be used to model an APT as a multi-stage attack on the network. 
This approach highlights the merits of breaking down APT multi-stage attacks into the stages relating to the adversarial tactics and identifying the specific techniques used by an attacker for each stage or tactic. 
Even if it is not possible to detect and mitigate all stages of the attack through a security risk assessment, it might be possible to detect enough stages to identify an APT and provide sufficient defences to prevent the attacker reaching their objective by eliminating critical paths.

In this section we demonstrate the use of the 5GCN knowledge base of adversarial techniques to model theoretical multi-stage attacks.


In the following examples we use threat intelligence from historic APT campaigns to support adversarial motivation and reproduce these using the newly identified 5GCN TTPs.

\subsection{Data Theft Scenario}
\begin{figure}
\centering{\includegraphics[width=0.75\textwidth]{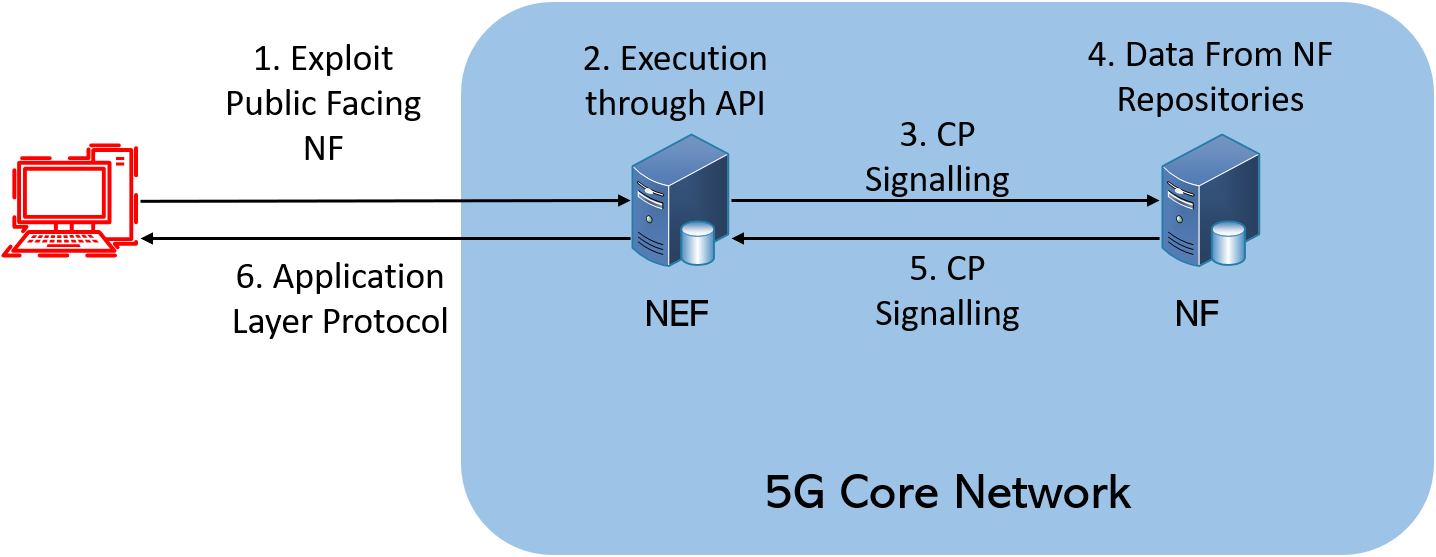}}
\caption{Scenario 1: A data theft scenario\label{DataTheftScenario}}
\end{figure}

A report published by Fireeye into APT41 includes threat intelligence about activities which targeted telecommunication networks \cite{dragon2020double}. 
It is claimed that the group targeted call record information as part of wider intelligence gathering efforts. 
The following attack scenario, illustrated in Figure \ref{DataTheftScenario}, considers the objective of the APT to extract confidential data from the 5GCN.

\begin{figure*}
\centering{\includegraphics[width=1\textwidth]{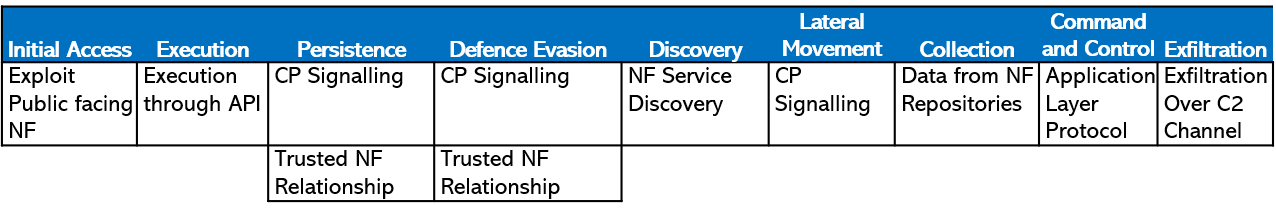}}
\caption{Scenario 1 TTPs \label{DataTheftTTP}}
\end{figure*}

In the beginning the attacker targets a public facing NF to gain initial access through an API exploit to compromise the NF. 
With this C2 channel established between the attacker and compromised NF, the attacker is able to discover the available NFs registered within the SBA.
CP signalling is used to request data from a target NF, the attacker uses CP signalling along with the trusted relationships between the registered NFs within the SBA to remain persistent and undetected from network defences. 
On receipt of the service request, the target NF accesses the requested data from its data repository and returns it to the compromised NF which requested it. 
Following the data collection stage, it is exfiltrated out of the 5GCN using the application layer protocol to conceal the contents.
In Figure \ref{DataTheftTTP} the attack vectors relating to this scenario are shown. 


\subsection{MANO Service Abuse}
\begin{figure}
\centering{\includegraphics[width=0.75\textwidth]{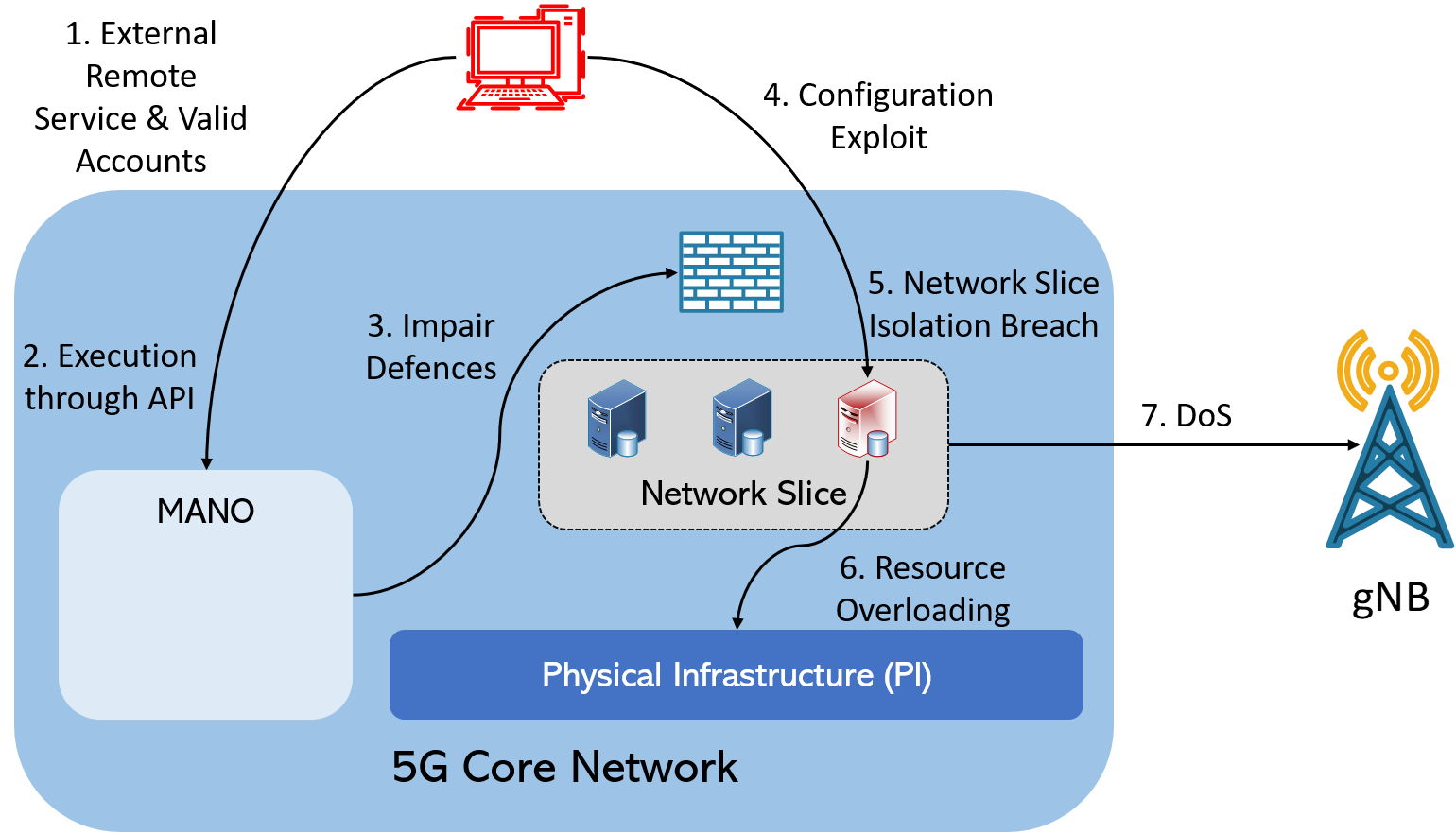}}
\caption{Scenario 2: MANO Attack Scenario\label{MANOScenario}}
\end{figure}

In order to provide scalability and flexibility in service provision, the 5GCN will use SDN and NFV providing a reprogammable and dynamic network architecture. 
To support on demand services and different use case requirements, reconfiguration of the 5GCN components is likely to be a commonly repeated task.  
Misconfiguration of the virtual infrastructure can pose a significant threat to security, in the cloud environment, whether malicious or unintentional \cite{iqbal2016cloud}. 
In this scenario abuse of the MANO is considered as shown in Figure \ref{MANOScenario}.

\begin{figure*}
\centering
    \includegraphics[width=1\textwidth]{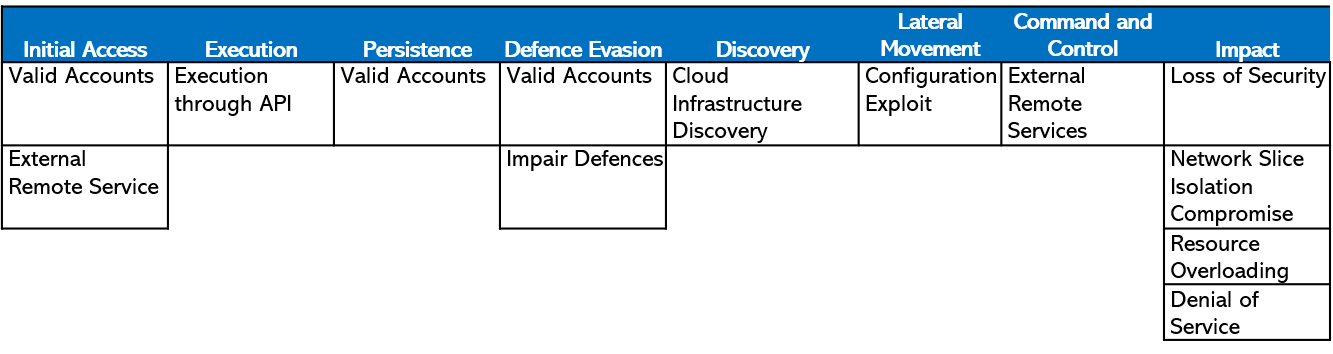}
	\caption{Scenario 2 TTPs} \label{MANOTTP}
\end{figure*}

The MANO component is targeted for initial access to the 5GCN through the external remote service and use of valid user credentials. 
Using the management API the firewall settings are modified allowing the attacker to bypass security controls the impact of this step is a Network Slice isolation. 
With the NF now exposed to the outside attacker, a DoS is launched leading to the underlying physical resources being exhausted. 
The end result is a DoS on the User Equipment (UE) being served by the targeted Network Slice.
In Figure \ref{MANOTTP} the adversarial techniques for each stage of the attack are mapped to the 5GCN TTP matrix.

\section{Conclusion}
\label{sec:concl}
In this paper, we identify the MITRE ATT\&CK as a suitable framework for threat modelling in the 5GCN lending to the dynamic and elastic properties of the network architecture. 
Through evaluation of 5GCN threat assessments, we identify adversarial techniques applicable to 5G networks to extend the existing ATT\&CK framework knowledge base, provide a mapping to the tactical stages of an APT lifecycle and demonstrate a practical implementation for modelling multi-stage attack scenarios. 
Our approach towards identification of adversarial techniques and mapping those to the tactical stages of an APT campaign and 5GCN infrastructure, adopts a preemptive methodology to allow for application of the well established MITRE ATT\&CK framework for threat modelling in the 5GCN and to support future 5GCN cyber risk assessments. 

\subsection{Limitations}
The roll-out of 5G core networks is anticipated over the coming years with the specification still under development. 
In the absence of real world threat intelligence reports for populating a 5G knowledge base which extends MITRE ATT\&CK, we have adopted a forward thinking approach to producing one. 
Our work is based on relevant threat assessments produced by academics and industry but lacks the real world analytics at this stage to support the extension to ATT\&CK.

Currently our proposed knowledge base addresses the identification of potential security risks in the form of TTPs but does not identify suitable detection and mitigation techniques as yet. 
This in part is due to the lack of available data relating to APT campaigns against 5G networks but also because many of the new technologies are yet to be deployed.

The significance of security risk assessment is understanding the security risk to networks with consideration of likelihood and impact of given attack scenarios. 
There are several factors that contribute to the process including detection/mitigation capabilities, security controls which are in place, the capability of the adversary and network configuration. 
These challenges are yet to be addressed and may not be until real world deployments of 5G networks are rolled out and analytics relating to attacks are available. 

\subsection{Future Work}
To further support and extend the proposed approach, the following future work could involve:
\begin{itemize}
    \item The simulation of identified 5GCN attack scenarios with generation of data sets providing further analysis of the techniques which have been identified. 
    In the absence of real world data this can support the inclusion of identified TTPs.
    \item Identification of suitable detection and mitigation capability requirements for the 5GCN adversarial techniques identified in this work.  
    \item A cyber security risk assessment incorporating the newly identified 5GCN TTPs to provide a holistic evaluation of security risk to the 5GCN.
\end{itemize}
Future work may include the creation of a framework to model 5GCN using requirements similar to our previous work \cite{mavropoulos2019apparatus,mavropoulos2017conceptual}.
We also plan to propose a game-theoretic framework, which will study the interactions between an agent that defends the 5G infrastructure and an attacking agent, e.g.\cite{rontidis2015game,panaousis2017game}.

\section*{Acknowledgement}
This work was partly funded by a UK Government PhD Studentship Scheme.

\bibliographystyle{unsrtnat}
\bibliography{references}  






\end{document}